\documentclass[twocolumn,showpacs,preprintnumbers,amsmath,amssymb]{revtex4}

\usepackage{graphicx}
\usepackage{dcolumn}
\usepackage{bm}

\newcommand{\mb}[1]{ \mbox{\boldmath$#1$} }
\newcommand{\ds}{\displaystyle}
\newcommand{\beq}{\begin{eqnarray}}
\newcommand{\eeq}{\end{eqnarray}}
\newcommand{\beqq}{\begin{eqnarray*}}
\newcommand{\eeqq}{\end{eqnarray*}}
\newcommand{\p}{\partial}

\newcommand{\x}{\mbox{\boldmath$x$}}
\newcommand{\w}{\mbox{\boldmath$w$}}

\font\bb=msbm10 at 12pt

\def\rR{\hbox{\bb R}}

\begin{document}
\title
{LANGEVIN TRAJECTORIES BETWEEN FIXED CONCENTRATIONS\\}

\author{B. Nadler}
\email{boaz.nadler@yale.edu}
 \affiliation{Department of
Mathematics, Yale University, 10 Hillhouse Ave. P.O. Box 208283, New Haven, CT 06520-8283}
 \author{Z. Schuss}
\email{schuss@post.tau.ac.il}
\author{A. Singer}
\email{amits@post.tau.ac.il} \affiliation{Department of Applied
Mathematics, Tel-Aviv University, Ramat-Aviv, 69978 Tel-Aviv,
Israel}

\date{\today}

\begin{abstract}
We consider the trajectories of particles diffusing between two
infinite baths of fixed concentrations connected by a channel, e.g.
a protein channel of a biological membrane. The steady state influx
and efflux of Langevin trajectories at the boundaries of a finite
volume containing the channel and parts of the two baths is
replicated by termination of outgoing trajectories and injection
according to a residual phase space density. We present a simulation
scheme that maintains averaged fixed concentrations without creating
spurious boundary layers, consistent with the assumed physics.
\end{abstract}


\maketitle

\section{\label{sec:intro}Introduction}
We consider particles that diffuse in a domain $\Omega$ connecting
two regions, where fixed, but possibly different concentrations are
maintained by connection to practically infinite reservoirs. This is
the situation in the diffusion of ions through a protein channel of
a biological membrane that separates two salt solutions of different
fixed concentrations \cite{Hille}.

Continuum theories of such diffusive systems describe the
concentration field by the Nernst-Planck equation (NPE) with fixed
boundary concentrations \cite{Hille}-\cite{SNE1}. 
On the other hand, the underlying microscopic theory of diffusion
describes the motion of the diffusing particles by Langevin's
equations \cite{BRR}, \cite{SNE1}-\cite{book}. This means that on a
microscopic scale there are fluctuations in the concentrations at
the boundaries. The question of the boundary behavior of the
Langevin trajectories (LT), corresponding to fixed boundary
concentrations, arises both in theory and in the practice of
particle simulations of diffusive motion \cite{Tildesley}-\cite{Im}.

 When the concentrations are maintained by connection to infinite
reservoirs, there are no physical sources and absorbers of
trajectories at any definite location in the reservoir or in
$\Omega$.  The boundaries in this setup can be chosen anywhere in
the reservoirs, where the average concentrations are effectively
fixed. Nothing unusual happens to the LT there. Upon reaching the
boundary they simply cross into the reservoir and may cross the
boundary back and forth any number of times. Limiting the system to
a finite region necessarily puts sources and absorbers at the
interfaces with the baths, as  described in \cite{Unidir}.

The boundary behavior of diffusing particles in a finite domain
$\Omega$ has been studied in various cases, including absorbing,
reflecting, sticky boundaries, and many other modes of boundary
behavior \cite{Mandl}, \cite{Karlin}. In \cite{Schumaker} a sequence
of Markovian jump processes is constructed such that their
transition probability densities converge to the solution of the
Nernst-Planck equation with given boundary conditions, including
fixed concentrations and sticky boundaries. Brownian dynamics
simulations with different boundary protocols seem to indicate that
density fluctuations near the channels are independent of the
boundary conditions, if the boundaries are moved sufficiently far
away from the channel \cite{Corry2002}. However, as shown in
\cite{PRE}, many boundary protocols for maintaining fixed
concentrations lead to the formation of spurious boundary layers,
which in the case of charged particles may produce large long range
fluctuations in the electric field that spread throughout the entire
simulation volume $\Omega$. The analytic structure of these boundary
layers was determined in \cite{Marshall, Hagan}, following several
numerical investigations (e.g, \cite{Titulaer81}).

It seems that the boundary behavior of LT of particles diffusing
between fixed concentrations has not been described mathematically
in an adequate way. From the theoretical point of view, the
absence of a rigorous mathematical theory of the boundary behavior
of LT diffusing between fixed concentrations, based on the
physical theory of the Brownian motion, is a serious lacuna in
classical physics.

It is the purpose of this letter to analyze the boundary behavior of
LT between fixed concentrations and to design a Langevin simulation
that does not form spurious boundary layers. We find the joint
probability density function of the velocities and locations, where
new simulated LT are injected into a given simulation volume, while
maintaining the fixed concentrations. As the time step decreases the
simulated density converges to the solution of the Fokker-Planck
equation (FPE) with the imposed boundary conditions without forming
boundary layers.

\section{ Trajectories, fluxes, and boundary concentrations}

We assume fixed concentrations $C_L$ and $C_R$ on the left and right
interfaces between $\Omega$ and the baths $B$, respectively, with
all other boundaries of $\Omega$ being impermeable walls, where the
normal particle flux vanishes. We assume that the particles interact
only with a mean field, whose potential is $\Phi(\x)$, so the
diffusive motion of a particle in the channel and in the reservoirs
is described by the Langevin equation (LE)
\begin{eqnarray}
\ddot{\x}+\gamma(\x)\dot{\x}+ \nabla_{\x}\Phi(\x)&=&
\sqrt{2\gamma(\x)\varepsilon}\,\dot
{\w}\nonumber\\
&&\label{LEs}\\
\x(0)=\x_0,\quad \mb{v}(0)&=&\mb{v}_0,\nonumber
\end{eqnarray}
where $\gamma(\x)$ is the (state-dependent) friction per unit mass,
$\varepsilon$ is a thermal factor, and $\dot{\w}$ is a vector of
standard independent Gaussian $\delta$-correlated white noises
\cite{book}.

The probability density function (pdf) of finding the trajectory
of the diffusing particle at location $\x$ with velocity $\mb{v}$
at time $t$, given its initial position, satisfies the
Fokker-Planck equation (FPE) in the bath and in the reservoirs,
\begin{eqnarray}
\frac{\p p}{\p
t}&=&-\mbox{\boldmath$v$}\cdot\nabla_{\mbox{\boldmath$x$}}\,p+
\gamma(\mbox{\boldmath$x$})\varepsilon\Delta_{\mbox{\boldmath$v$}}p
\label{FPE} \\
&& +\nabla_{\mbox{\boldmath$v$}}\cdot
\Big[\gamma(\mbox{\boldmath$x$}) \mb{v}+
\nabla_{\mb{x}}\Phi(\mbox{\boldmath$x$})\Big]
p,\nonumber\\
p(\x,\mb{v},0\,|\,\x_0,\mb{v}_0)&=&\delta(\x-\x_0,\mb{v}-\mb{v}_0)\nonumber.
\end{eqnarray}
In the Smoluchowski limit of large friction the stationary solution
of (\ref{FPE}) admits the form \cite{EKS}
\begin{equation}
p(\mb{x},{\mb v})=\frac{e^{-|\mb{v}|^2/2\varepsilon }}{(2\pi
\varepsilon)^{3/2}} \left\{p(\mb{x})+\frac{{\cal J}(\mb{x})\cdot
\mb{v}}{\varepsilon }+O\left( \frac{1}{\gamma ^{2}}\right)
\right\} \label{fluxform}
\end{equation}%
where the flux density vector ${\cal J}(\mb{x})$ is given by
\begin{equation}
{\cal J}(\mb{x})=-\frac{1}{\gamma \left( \mb{x}\right) }\bigg\{
\varepsilon \nabla p(\mb{x})+p(\mb{x})\,\nabla\Phi(\mb{x})
\bigg\}+O\left( \frac{1}{\gamma ^{2}}\right) ,  \nonumber
\end{equation}%
and $p(\mb{x})$ satisfies
 \beqq
-\nabla\cdot{\cal
J}(\mb{x})=\nabla\cdot\frac{1}{\gamma\left(\mb{x}\right)}\bigg\{
\varepsilon\nabla
p(\mb{x})+p(\mb{x})\,\nabla\Phi\left(\mb{x}\right)\bigg\}=0.
    \label{SNPE}
 \eeqq
In one dimension, the stationary pdfs of velocities of the particles
crossing the interface into the given volume are
\begin{eqnarray}
p_L(v)&\sim& \frac{\displaystyle\frac{e^{-v^{2}/2\varepsilon
}}{\sqrt{2\pi\varepsilon }} \left\{ 1+\displaystyle\frac{{\cal
J}v}{\varepsilon C_L}\right\} }{\displaystyle\frac{1}{2}
+\displaystyle\frac{{\cal J}}{C_L\sqrt{2\pi\varepsilon }}}\quad
\mbox{for}\quad v>0, \nonumber \\
&&\label{pvR}\\
p_{R}(v)&\sim& \frac{\displaystyle\frac{e^{-v^{2}/2\varepsilon
}}{\sqrt{2\pi\varepsilon }} \left\{ 1-\displaystyle\frac{{\cal
J}v}{\varepsilon C_R}\right\} }{\displaystyle\frac{1}{2}
+\displaystyle\frac{{\cal J}}{C_R\sqrt{2\pi\varepsilon }}}\quad
\mbox{for}\quad v<0,\nonumber
\end{eqnarray}
where ${\cal J}$ is the net probability flux through the channel.
The source strengths (unidirectional fluxes at the interfaces) are
given by \cite{EKS}
 \beq
J_{L}&=&\sqrt{\frac\varepsilon{2\pi}}C_L-\frac{{\cal
J}}2+O\left(\frac{1}{\gamma^2}\right)\nonumber\\
&&\label{unismolRL}\\
J_{R}&=&\sqrt{\frac\varepsilon{2\pi}}C_R+\frac{{\cal J}}2
+O\left(\frac{1}{\gamma^2}\right).\nonumber
 \eeq

\section{Application to simulation}
Langevin simulations of ion permeation in a protein channel of a
biological membrane have to include a part of the surrounding bath,
because boundary conditions at the ends of the channel are unknown.
The boundary of the simulation has to be interfaced with the bath in
a manner that does not distort the physics. This means that new LT
have to be injected into the simulation at the correct rate and with
the correct distribution of displacement and velocity, for
otherwise, spurious boundary layers will form \cite{PRE}.

Consider a single simulated trajectory that jumps according to the
discretized LE (\ref{LEs})
 \beq
 \x(t+\Delta t)&=&\mb{v}(t)\Delta t,\quad
 \mb{v}(t+\Delta t)=\mb{v}(t)(1-\gamma\Delta
 t)\nonumber\\
 &-&\nabla_{\x}\Phi(\x(t))\Delta t+\sqrt{2\varepsilon\gamma}\,\Delta
 \mb{w}(t),\label{DLE}
 \eeq
where $\Delta w$ is normally distributed with zero mean and variance
$\Delta t$. The trajectory is terminated when it exits $\Omega$ for
the first time. The problem at hand is to determine an injection
scheme of new trajectories into $\Omega$ such that the interface
concentrations are maintained on the average at their nominal values
$C_L$ and $C_R$ and the simulated density profile satisfies
(\ref{fluxform}).

To be consistent with (\ref{fluxform}), the injection rate has to be
equal to the unidirectional flux at the boundary (\ref{unismolRL}).
New trajectories have to be injected with displacement and velocity
as though the simulation extends outside $\Omega$, consistently with
the scheme (\ref{DLE}), because the interface is a fictitious
boundary. The scheme (\ref{DLE}) can move the trajectory from the
bath $B$ into $\Omega$ from any point $\mb{\xi}\in B$ and with any
velocity $\mb{\eta}$. The probability that a trajectory, which is
moved with time step $\Delta t$ from the bath into $\Omega$, or from
$\Omega$ into the bath will land exactly on the boundary is zero. It
follows that the pdf of the point $(\x,\mb{v})$, where the
trajectory lands in $\Omega$ in one time step, at time  $t' =
t+\Delta t$, say, given that it started at a bath point
$(\mb{\xi},\mb{\eta})$ (in phase space) is, according to
(\ref{DLE}),
 \beq
&&\Pr\{\x(t')=\x,\mb{v}(t')=\mb{v}\,|\,\x(t)=\mb{\xi},\mb{v}(t)=\mb{\eta}\}\nonumber\\
 &&=\frac{\delta(\x-\mb{\xi}-\mb{\eta}\Delta
 t)}{(4\pi\varepsilon\gamma\Delta t)^{3/2}}\times \label{LanDyn} \\
&& \exp\left\{-\ds\frac{\left|\mb{v}-
 \mb{\eta}-(\gamma\mb{v}+\nabla\Phi(\mb{\xi}))\Delta
 t\right|^2}{4\varepsilon\gamma\Delta t}\right\}+o(\Delta t).\nonumber
 \eeq
The stationary pdf $p(\mb{\xi},\mb{\eta})$ of such a bath point is
given in (\ref{fluxform}). The conditional probability of such a
landing is
 \beq
&&\Pr\{\x,\mb{v}\,|\,\x \in\Omega,\mb{\xi}\in B\}=\label{residual}\\
&&\frac{\ds\int_{\rR^3}d\mb{\eta}\ds\int_{B}d\mb{\xi}
\Pr\{\mb{v}(t')=\mb{v},\x(t')=\x\,|\,\mb{\xi},\mb{\eta}\}
p(\mb{\xi},\mb{\eta})}{\Pr\{\x\in\Omega,\mb{\xi}\in B\}},\nonumber
 \eeq
where the denominator is a normalization constant such that
 \beqq
 \int_{\rR^3}d\mb{v}\int_{\Omega}d\x\,\Pr\{\x,\mb{v}\,|\,\x
 \in\Omega,\mb{\xi}\in\mbox{B}\}=1.
 \eeqq
Thus the first point of a new trajectory is chosen according to the
pdf (\ref{residual}) and the subsequent points are generated
according (\ref{DLE}), that is, with the transition pdf
(\ref{LanDyn}), until the trajectory leaves $\Omega$. By
construction, this scheme recovers the joint pdf (\ref{fluxform}) in
$\Omega$, so no spurious boundary layer is formed.

As an example, we consider a one-dimensional Langevin dynamics
simulation of diffusion of free particles between fixed
concentrations on a given interval. Assuming that in a channel of
length $L$
 \beqq
 \frac{(C_L-C_R)\sqrt{\varepsilon}}{\gamma L}\ll C_L,
 \eeqq
which means that $\gamma$ is sufficiently large, the flux term in
eq.(\ref{fluxform}) is negligible relative to the concentration
term. The concentration term is linear with slope ${\cal J}$ and
thus can be approximated by a constant, so that
$p(\xi)=p(0)+O\left(\gamma^{-1}\right)$ in the left bath. Actually,
the value of $p(0)\neq0$ is unimportant, because it cancels out in
the normalized pdf (\ref{residual}), which comes out to be
 \beq
\Pr\{x,v\,|\,x>0,\xi<0\} =
\frac{\exp\left\{-\ds\frac{v^2}{2\epsilon[1+(\gamma \Delta t)^2]}
\right\}}{2\epsilon \Delta t \sqrt{1+(\gamma \Delta
t)^2}}&& \nonumber \\
&&\label{residual1}\\ \times \mbox{erfc}
\left(\sqrt{\ds\frac{1+(\gamma \Delta t)^2}{4\epsilon \gamma
\Delta t}} \left(\frac{x}{\Delta t} - v\frac{1-\gamma \Delta
t}{1+(\gamma \Delta t)^2}\right) \right).&&\nonumber
 \eeq
In the limit $\Delta t\to0$ we obtain from
eq.(\ref{residual1})
 \beq
\Pr\{x,v\,|\,x>0,\xi<0\}
\to\frac{2\delta(x)H(v)}{\sqrt{2\pi\varepsilon}}e^{-v^2/2\varepsilon},\label{Delta}
  \eeq
where $H(v)$ is the Heaviside unit step function. This means that
with the said approximation, LT enter at $x=0$ with a Maxwellian
distribution of positive velocities. Without the approximation the
limiting distribution of velocities is (\ref{pvR}). Note, however,
that injecting trajectories by any Markovian scheme, with the
limiting distribution (\ref{Delta}) and with any time step $\Delta
t$, creates a boundary layer \cite{PRE}.

\begin{widetext}
\begin{figure}
\includegraphics{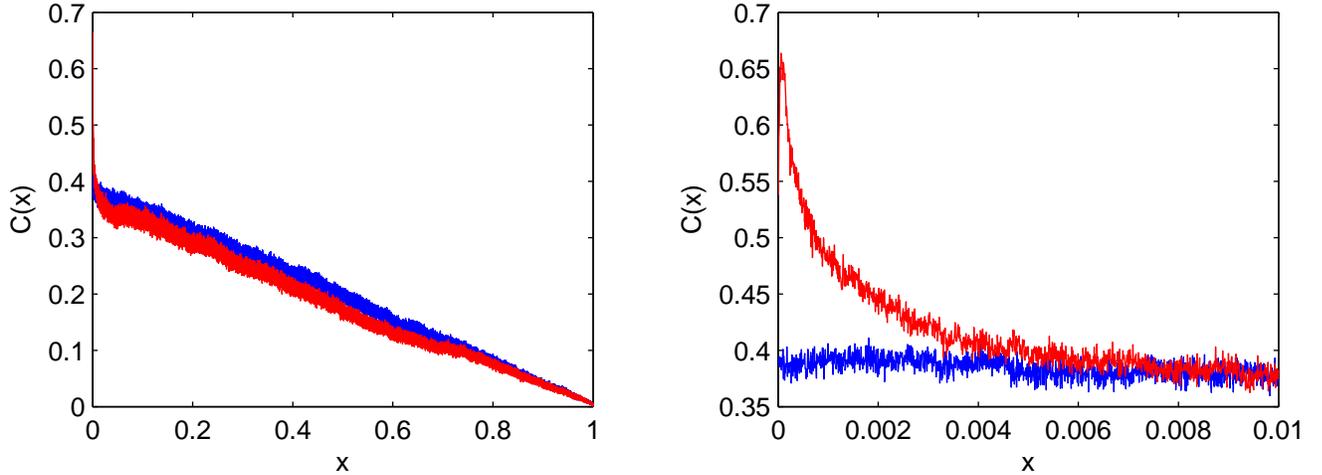}
\caption{Left panel: Concentration against displacement of a LD
simulation with injecting particles according to the residual
distribution (\ref{residual1}) (blue), and according to the
Maxwellian velocity distribution (\ref{Delta}) exactly at the
boundary (red). The two graphs are almost identical, except for a
small boundary layer near $x=0$ in red. Right panel: Zoom in of the
concentration profile in the boundary layer
$x<0.01=\sqrt{\epsilon}/\gamma$.} \label{f:both}.
\end{figure}
\end{widetext}

A LD simulation with $C_L\neq0,\ C_R=0$, and the parameters $
\gamma=100,\ \varepsilon=1,\ L=1,\ \Delta t=10^{-4}$ with 25000
trajectories, once with a Maxwellian distribution of velocities at
the boundary $x=0$ (red) and once with the pdf (\ref{residual1})
(blue) shows that a boundary layer is formed in the former, but not
in the latter (see Figure \ref{f:both}).

An alternative way to interpret eq.(\ref{residual1}) is to view the
simulation (\ref{DLE}) as a discrete time Markovian process
$(\x(t),\mb{v}(t))$ that never enters or exits $\Omega$ exactly at
the boundary. If, however, we run a simulation in which particles
are inserted at the boundary, the time of insertion has to be
random, rather than a lattice time $n\Delta t$. Thus the time of the
first jump from the boundary into the domain is the residual time
$\Delta t'$ between the moment of insertion and the next lattice
time $(n+1)\Delta t$. The probability density of jump size in both
variables has to be randomized with $\Delta t'$, with the result
(\ref{residual1}).

\begin{acknowledgments}
The authors are indebted to M. Schumaker for making a preprint of
his work \cite{Schumaker} available to them. This work was the
motivation for the present paper. This research was partially
supported by research grants from the US-Israel Binational Science
foundation, the Israel Science Foundation, and DARPA.
\end{acknowledgments}



\end{document}